\documentclass[twoside,slac_one]{revtex4}
\usepackage{graphicx}
\usepackage{fancyhdr}
\usepackage{amsmath} 
\usepackage{bm}
\usepackage{amsxtra}
\usepackage{amssymb}
\usepackage{amsthm}
\usepackage{latexsym}
\usepackage{lscape}

\newcommand{\met}{\mbox{${\not\!\!E_{\rm T}}$}}

\newcommand{\fb}{\ensuremath{\mathrm{fb}^{-1}}}
\newcommand{\KeV}{\ensuremath{\mathrm{Ke\kern -0.1em V}}}
\newcommand{\MeV}{\ensuremath{\mathrm{Me\kern -0.1em V}}}
\newcommand{\GeV}{\ensuremath{\mathrm{Ge\kern -0.1em V}}}
\newcommand{\TeV}{\ensuremath{\mathrm{Te\kern -0.1em V}}}
\newcommand{\ttbar}{\ensuremath{t\bar{t}}}
\newcommand{\mtop}{\ensuremath{M_{top}}}
	
\begin{document}

\title{Measurements of the $t\bar{t}$ Cross Section and the Top Quark Mass in the Hadronic $\tau$ + Jets Decay Channel at CDF}
%

\author{D. Hare\footnote{Speaker} and E. Halkiadakis}
\affiliation{Department of Physics and Astronomy, Rutgers University, New Brunswick, NJ, USA}
\author{On behalf of the CDF Collaboration}

\begin{abstract}
 We present the first exclusive observation of the $\ttbar \rightarrow$ hadronic $\tau$ + jets decay channel. Using these events from 1.96 $\TeV$ $p\bar{p}$ collisions at CDF, we measure the $t\bar{t}$ cross section as well as the top quark mass. Events require a single hadronic $\tau$, large missing transverse energy, and exactly 4 jets of which at least one must be tagged as a $b$ jet. The cross section measurement is extracted from a Poisson likelihood function based on the observed number of events and the predicted number of signal and background events for a given $t\bar{t}$ cross section. The mass is extracted from a likelihood fit based on per-event probabilities calculated from leading-order signal ($t\bar{t}$) and background ($W$+jets) matrix elements.
\end{abstract}

\maketitle

\thispagestyle{fancy}

\graphicspath{{figures/}}
\section{Introduction}
We present the first exclusive observation of $\ttbar \rightarrow$ hadronic ~$\tau$ + jets events. With these events, we measure the $t\bar{t}$ production cross section in $\bar{p}p$ collisions at $\sqrt{s}=1.96 ~\TeV$ with the CDF detector \cite{cdfdet} at the Tevatron at Fermilab, as well as the first direct measurement of the top quark mass in $\tau$ + jets events.  These measurements provide important tests of lepton universality and probe the top quark properties in a relatively unexplored channel which may be sensitive to new physics.  Additionally, they are good examples of physics measurements performed with $\tau$ leptons in high jet multiplicity environments.

\section{Selection and Background Estimation}
 This analysis uses a dataset with a total integrated luminosity of 2.2 $\fb$ collected with the CDF detector between February 2002 and August 2007.  The data is selected using a multi-jet trigger which requires at least four jets each with a calorimeter cluster with transverse energy ($E_T$, where transverse refers to being perpendicular to the beamline) $> 15 ~\GeV$ and a total sum $E_T$ of all reconstructed jets $>175 ~\GeV$.  To these events, we apply selection criteria which require 4 jets with $E_T> 20 ~\GeV$, missing $E_T$ ($\met$) $> 20 ~\GeV$, and a hadronically decaying $\tau$ lepton with $E_T> 25 ~\GeV$.  Additionally, one of the 4 jets must be identified as coming from a $b$ quark ($b$-tagging) \cite{secvtx}.  Since our signal process gives a single $\tau$ lepton, we veto any event with an identified electron or muon.  Hadronically decaying $\tau$'s appear as narrow jets with an odd number of charged tracks and low $\pi^0$ multiplicity. They are selected using similar requirements as described in \cite{tauSel}, except we require both 1 and 3 prong $\tau$'s to have visible $E_T$ of at least $25 ~\GeV$ and a visible mass less than $1.8 ~\GeV$.  We also place no explicit requirement on the transverse energy of the $\pi^0$'s in the isolation region, but we do require that calorimeter energy in the isolation region be less than 10\% of the $\tau$ energy.
  	
\subsection{\label{sec:NN} Neural Network for QCD Multijets Removal}
\begin{figure}[htbp]
	\begin{center}
	\includegraphics[width=6cm]{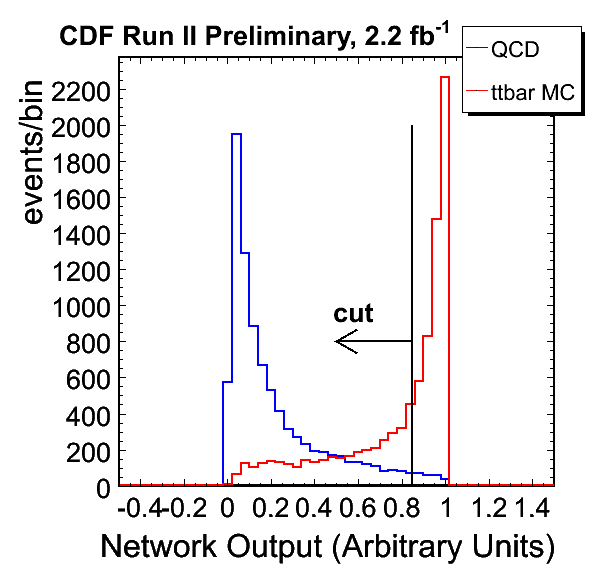}  
	\caption{NN output distribution for signal ($\ttbar$ in red) and background (QCD multijets in blue) events.  The NN is trained with $\ttbar$ events given an output value of 1 and background events given an output value of 0. We select events with a NN output value $> 0.85$.\label{fig:NNoutput}}
	\end{center}
	\end{figure}

The dominant background for this analysis is high jet multiplicity QCD events with one of the jets faking the signature of a $\tau$ lepton. To further reduce the QCD multijets background, we developed an artificial neural network (NN) to distinguish between true $t\bar{t} \rightarrow \tau$ + jets events and QCD multijets events.  First, we create a sample of QCD multijets from data by selecting events with a $\tau$ with no track isolation requirement.  The NN is trained to distinguish between these selected QCD multijets events and $t\bar{t}$ events generated with the {\sc{Pythia}} MC generator \cite{pythia} where the $\tau$ decay is handled by the {\sc{Tauola}} package \cite{tauola} to properly account for the $\tau$ polarization.  We use 8 variables to train the NN: $\met$, lead jet E$_T$, sum E$_T$ of the jets and $\tau$ lepton, sum E$_T$ of the two lowest E$_T$ jets and the $\tau$ lepton, sum E$_T$ of the two highest E$_T$ jets, transverse momentum of the $W$ which decays to a $\tau$ lepton, average $\eta$-moment of all jets not identified as coming from a $b$ quark, and the lowest ratio of dijet mass to trijet mass for any possible triplet of jets.  After training the NN, we find it provides good separation between QCD multijets and $t\bar{t}$ events as can be seen in Fig. \ref{fig:NNoutput}, and we find the optimal signal significance is achieved by removing all events which return a NN output below 0.85.  Before the NN requirement, we select 162 $\tau$ + jets events which are largely QCD multijets events as is shown in Fig. \ref{fig:tagFit}.  After the NN selection is applied, we find 41 events of which we expect roughly 18 QCD multijets events and 18 $\ttbar$ events.  From MC studies, we estimate 76\% of the selected $\ttbar$ events are hadronic $\tau + jets$ decays.  The majority of the remaining $\ttbar$ events come from all-hadronic $\ttbar$ decays with less than 7\% contamination from $\ttbar \rightarrow e$ + jets.
	\subsection{\label{sec:M2} Background Estimation}
Due to the difficulty in MC modeling of QCD multijets events, $b$ quark tagging algorithms, and the production of heavy flavor quarks in association with $W$ bosons, we use a data-driven approach to estimate the background contribution similar to that described in \cite{Aaltonen:2010ic}. First, we calculate the contributions from electroweak background processes which have a minimal contribution to the final total (diboson, single top quark production, and $Z$ + jets events), as well as the $\ttbar$ signal contribution, by using the theoretical cross section for each process along with the acceptance from MC simulation and the total integrated luminosity.  

With these contributions known, we evaluate the contributions for QCD multijets and $W$ + jets events by fitting the shape of the NN output distribution for each component (including the fixed contributions already calculated) to the data before the NN selection and $b$-tagging requirements are applied.  We fit these distributions using a binned Poisson likelihood. From this fit, we evaluate the percentage of the data events above the 0.85 NN output value which are coming from QCD multijets.  Any remaining events are assumed to come from $W$ + jets processes.  

\begin{figure}[htbp]
	\begin{center}
	\includegraphics[width=6cm]{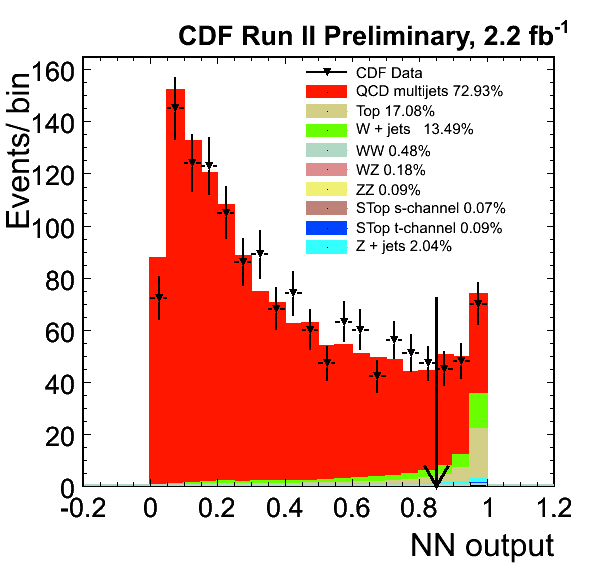} \\
	\includegraphics[width=6cm]{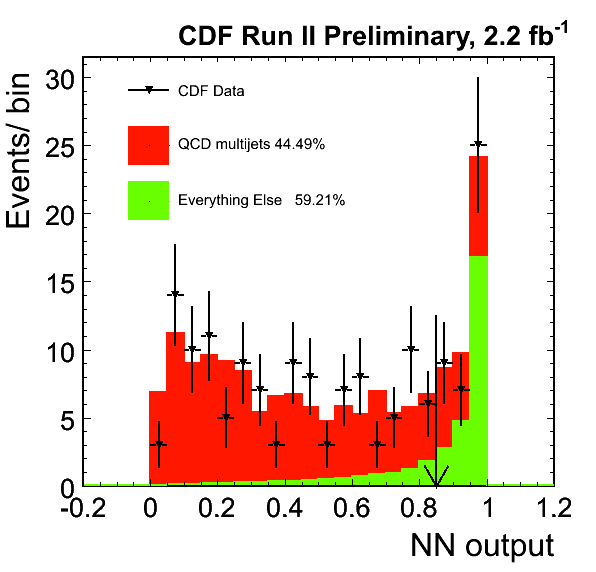}  
	\caption{Fit to the NN output shape beofre (top) and after (bottom) applying the $b$-tagging requirement.  The percentages listed in the legend are the percentage of events from each source contributing to the data sample after applying the NN selection requirement (shown with an arrow).\label{fig:tagFit}}
	\end{center}
	\end{figure}

We next apply $b$-tagging efficiencies to sources except the data-based QCD multijet events to estimate the resulting contribution from each source after the $b$-tagging requirement is applied. The $W$ + jets events are divided up into contributions from $W$ + light flavor and $W$ + heavy flavor ($W+bb$, $W+cc$, and $W+c$).  We then fit the resulting NN output shapes along with the QCD multijet NN output shape to the data after the $b$-tagging requirement is applied to calculate the contribution from QCD multijet events. The fits before and after the $b$-tagging requirement are shown in Fig. \ref{fig:tagFit}. The contribution for each process assuming a top pair cross section of 7.4 pb and a top quark mass of 172.5 $\GeV$ is shown in Tab. \ref{tab:tau_m2}.

	\begin{table}
	\begin{center}
	\begin{tabular}{|l|c|}
	\hline
	Source & Number of Events \\
	\hline
	Diboson           & $  0.19 \pm  0.01$  \\
	Single top        & $  0.16 \pm  0.01$  \\
	$Zbb$             & $  0.29 \pm  0.04$  \\
	$Wbb$             & $  0.57 \pm  0.47$  \\
	$Wcc$             & $  0.34 \pm  0.28$  \\
	$Wc$              & $  0.15 \pm  0.13$  \\
	$W$+lf            & $  0.46 \pm  0.60$  \\
	QCD multijets     & $ 18.24 \pm  4.10$  \\
	\hline
	Total Bkgd      & $ 20.40 \pm 4.18 $ \\
	\hline
	Top             & $ 18.17 \pm  2.79$  \\
	\hline
	Total Predicted & $ 38.57 \pm 5.05 $  \\
	\hline
	Observed        & 41 \\
	\hline
	\end{tabular}
	\caption{Predicted number of selected $\tau$ + jet events from each considered process after applying a NN selection value of 0.85 assuming a $t\bar{t}$ pair production cross section of 7.4 pb and top quark mass of 172.5 $\GeV$.  We expect roughly 39 events compared to the observed 41 with nearly half of the events coming from $t\bar{t} \rightarrow \tau$ + jets and half from QCD multijet production. The uncertainties given are a combination of statistical uncertainties and the selection efficiency uncertainties.  The QCD multijet uncertainty includes the systematic uncertainty on the fraction of QCD multijet events.\label{tab:tau_m2}}
	\end{center}
	\end{table}
	
	\section{Top Anti-Top Production Cross Section Measurement}
	Generally, cross sections are measured as:
\begin{equation}
\sigma = \frac{N_{data} - N_{bkgd}}{Acc \cdot \epsilon \cdot \cal{L}},
\label{form:xsec}
\end{equation}
where $N_{data}$ and $N_{bkgd}$ are the number of events observed in the data and the number of predicted background events, respectively.  The kinematic acceptance for the process being observed (for the case of $\sigma_{\ttbar}$, we measure here the acceptance for $p\bar{p} \rightarrow \ttbar \rightarrow \tau$ + jets is $Acc$, $\epsilon$ is the product of all geometrical and kinematic event selection efficiencies corrected for by data/MC scale factors (SF) when relevant, and $\cal{L}$ is the total integrated luminosity of the data.

\begin{figure}[htbp]
	\begin{center}
	\includegraphics[width=6cm]{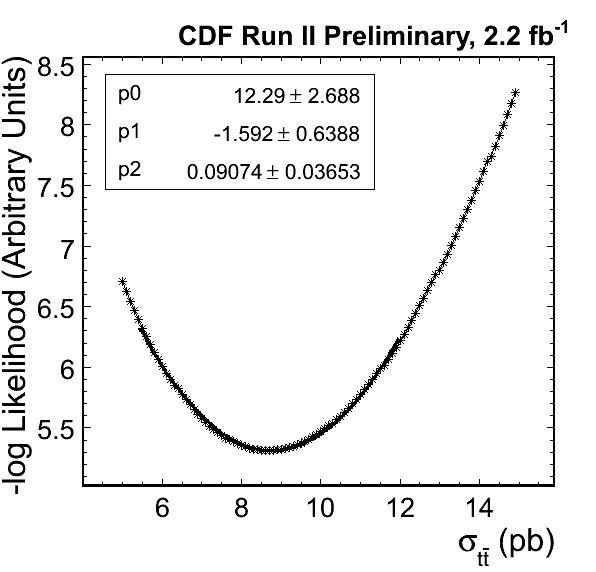}  
	\caption{The function $-2\cdot ln L$ versus input $\sigma_{t\bar{t}}$ as defined in Eq. \ref{eq:xsecLike}.  The solid line is the fit of a second order polynomial used to extract the central value and statistical uncertainty.\label{fig:xsecLike}}
	\end{center}
	\end{figure}

However, in this analysis, $N_{bkgd}$ is a function of the $\ttbar$ cross section ($\sigma_{\ttbar}$), as described in Sec. \ref{sec:M2}, and as a result, we cannot simply use Eq. \ref{form:xsec} to measure $\sigma_{\ttbar}$.  Instead, we build a likelihood function based on the Poisson probability distribution comparing the number of observed events and the predicted number of events.  We then minimize the negative log of this likelihood function written as:
\begin{equation}
-2 \cdot ln L = -2 \cdot \left( N_{data} \cdot ln \left( \sigma_{t\bar{t}} \cdot D + N_{b}(\sigma_{t\bar{t}}) \right) - ln \left( N_{data}! \right) - \left(\sigma_{t\bar{t}} \cdot D + N_{b}(\sigma_{t\bar{t}}) \right) \right),
\label{eq:xsecLike}
\end{equation}
where D is the denominator of Eq. \ref{form:xsec} and $N_{b}(\sigma_{t\bar{t}})$ ~is the number of events from the background prediction for a given $\sigma_{t\bar{t}}$. The result, shown in Fig. \ref{fig:xsecLike}, is fit with a $2^{nd}$ order polynomial which is used to extract the central value and statistical uncertainty.  We find the cross section to be $8.7 \pm 3.3 ~(\mathrm{stat.}) ~\mathrm{pb}$.
		
\subsection{\label{sec:xsecSyst} Top Pair Production Cross Section Systematic Uncertainties and Result}
To measure the systematic uncertainty on the $\sigma_{\ttbar}$ measurement we consider effects on the acceptance, selection efficiencies, background estimate, and luminosity.

For acceptance effects, we consider uncertainties on the jet energy scale (JES), initial and final state radiation (ISR, FSR), color reconnection,
parton showering, and parton distribution functions (PDF).  The jet energy measured by the calorimeter is subject to several correction functions each with an associated systematic uncertainty \cite{jes}.  To measure this uncertainty on the cross section, we shift the JES accordingly in the
MC and re-measure the cross section.  Changes in the amount of initial and final state radiation would change our acceptance, therefore, we model this effect using {\sc{pythia}} MC models with increased and decreased radiation \cite{ifsr}.  Similarly, we consider acceptance shifts from using models with and without color reconnection effects \cite{cr}.  We consider a 6\% systematic uncertainty for differences in parton showering models from different MC generators.  This number is taken from the difference in $\ttbar \rightarrow \tau$ + jets acceptance from MC generated with {\sc{Pythia}} and {\sc{Herwig}} \cite{herwig} after requiring the selected $\tau$ to be matched to a generated $\tau$ in the MC.  This requirement is applied because we find that {\sc{Herwig}} jets fake $\tau$'s at a rate higher than {\sc{Pythia}} jets, and our studies show that {\sc{Pythia}} better models the observed $\tau$ fakes in the data. Finally, we consider changes in acceptance by varying the eigenvectors of CTEQ6M \cite{cteq6} PDF's.

We consider systematic uncertainties on the efficiency measurements on the $b$-tagging, mistag matrix, lepton identification, and trigger.
Each of these uncertainties are evaluated by re-measuring the cross section with the appropriate efficiency or scale factor adjusted by its
systematic uncertainty.  Due to inefficiencies in modeling $b$-tagging in the MC, we use a tagging scale factor \cite{tagSF} on MC jets matched to heavy flavor to account for the $b$-tagging requirement.  Similarly, due to the poor modeling of mistags in MC, we use a data-based parameterization to model the mistagging of MC jets from light flavor \cite{tagSF}. We consider similar shifts for the lepton identification \cite{tauSF} and trigger efficiency \cite{note9954} scale factors.

The background systematics come from the $W$ + heavy flavor ``K-factor'' uncertainty and the QCD multijets contribution. For the number of events predicted for $W$ + heavy flavor processes, a data/MC scale factor called the ``K-factor'' is used to correct for the fraction of $W$ + heavy flavor events observed in the MC \cite{Aaltonen:2010ic}.  To measure the uncertainty from the K-factor, we shift it within its errors and take the difference in the cross section measurement as the uncertainty. Since QCD multijets events make up nearly 50\% of our accepted events, the contribution from QCD multijets events is our largest systematic uncertainty.  To measure this uncertainty, we select QCD multijets events from the data without the $\met$ and NN selection requirements.  We then compare the NN output distribution of these events to that of data events with the same requirements removed both before and after the $b$-tagging requirement.  The selection of data events without a $\met$ requirement is dominated by QCD multijets events below a NN value of 0.7, so we can fit the comparison between these distributions with a function to build a reweighting scheme for the QCD multijet distribution.  By shifting this fit function within its uncertainty, we define a 1$\sigma$ uncertainty on the shape of the QCD multijets distribution.  We then reweight the QCD multijets events and re-measure the cross section to measure the systematic uncertainty from the multijets contribution.  

Finally, we consider a 6\% uncertainty on the luminosity measurement from the detector accuracy and the uncertainty on the theoretical cross section for inelastic $p\bar{p}$ collisions \cite{bib:lumi}.

A summary of the contribution from each systematic source is given in Tab. \ref{tab:systematics}.

We measure $\sigma_{\ttbar}$ assuming a top quark mass of 172.5 $\GeV$ to be $8.7 \pm 3.3 ~(\mathrm{stat.}) \pm 2.2 ~(\mathrm{syst.}) ~\mathrm{pb}$ which is in good agreement with the most recent CDF combination of $7.5 \pm 0.5$ pb \cite{CDFxsec}.
	
		\begin{table}
	\begin{center}
	\begin{tabular}{|l|c|c|}
	\hline
	Systematic          & $\delta \sigma$ (pb)& $\delta \sigma / \sigma$ (\%)\\
	\hline
	Jet Energy Scale   & $ 0.6 $ & 6.9 \\
	IFSR               & $ 0.5 $ & 5.7 \\
	Color Reconnection & $ 0.4 $ & 4.6 \\
	Tagging            & $ 0.4 $ & 4.6 \\
	Mistag Matrix      & $ 0.1 $ & 1.1 \\
	QCD Fraction       & $ 1.8 $ & 20.5\\
	K-Factor           & $ 0.1 $ & 1.1 \\
	Parton Showering   & $ 0.5 $ & 6.0 \\
	Lepton ID          & $ 0.2 $ & 2.3 \\
	Trigger Efficiency & $ 0.1 $ & 1.1 \\
	PDF                & $ 0.5 $ & 5.7 \\
	Luminosity         & $ 0.5 $ & 6.0 \\
	\hline
	Total         & $ 2.2 $ & 25.0 \\
	\hline
  \end{tabular}
	\caption{Systematic uncertainties for the $t\bar{t}$ cross section measurement in the $\tau$ + jets decay channel.  The uncertainties are given as well as the fractional uncertainty.\label{tab:systematics}}
	\end{center}
	\end{table}

\section{Top Quark Mass Measurement}
The top quark mass measurement is a Matrix Element style analysis.  The mass is extracted from a likelihood function based on signal and background probabilities for each event.  These probabilities, described in Sec. \ref{sec:topLike}, are calculated from the differential cross section for $\ttbar$ and $W$ + 4 parton production.  The $\tau$ lepton decay adds extra complication to the mass measurement because its decay introduces a second $\nu$ into the event.  To account for this, we develop a new method to reconstruct the $\nu$ from the $\tau$ decay which allows us to reconstruct the original $\tau$ lepton which is described next.

\subsection{Collinear $\nu$ Approximation}
The missing energy from the $\nu$ from the $\tau$ lepton decay complicates the measurement of the top quark mass in the hadronic $\tau$ + jets decay channel.  For this analysis, we developed a new method to reconstruct this $\nu$'s 4-momentum which in turn allows us to reconstruct the 4-momentum of the $\tau$ lepton before it decays.  We assume that the $\nu$ from the $\tau$ decay is nearly collinear with the hadronic components of the $\tau$ decay within 0.1 radians in $\theta$ and $\phi$. Additionally, we assume that the $\phi$ angle of the $\nu$ from the $W$ boson is within 1 radian of the $\met ~\phi$.  These assumptions come from studying MC simulation of $\ttbar \rightarrow \tau$ + jets events.  We introduce a 4 dimensional scan across the angles of both $\nu$'s.  Assuming the $\nu$ mass is negligible and that the $W$ and $\tau$ have masses of $80.4 ~\GeV$ and $1.8 ~\GeV$, respectively, we can completely solve for the 4-vector of each $\nu$ for any set of angles from the scan.  We then use the $\nu$ solutions to predict the $\met$ in the event and compare it to the event's measured $\met$ with a Gaussian probability function.  We chose the set of angles which returns the greatest probability based on the $\met$ comparison.  This method accurately reconstructs the 4-momentum of the $\nu$ from the $\tau$ lepton decay, but it does not perform as well with the 4-momentum of the $\nu$ from the $W$ decay.  We use the result to reconstruct the original $\tau$ lepton in the event.  Meanwhile, the $\nu$ from the $W$ decay is reconstructed in the Matrix Element method as described in \cite{Abulencia:2007br} and below.

\subsection{Top Quark Mass Likelihood Function}
\label{sec:topLike}
The top quark mass ($\mtop$) measurement is derived from a likelihood function based on signal and background probabilities for each event.  The
method uses a similar approach as a previous measurement in the electron and muon + jets decay channels \cite{Abulencia:2007br}.  The
signal probability is based on a $t\bar{t}$ leading-order matrix element which assumes $q\bar{q}$ production \cite{Mahlon:1995zn} and is calculated over 31 input mass values ranging from 145 to 205 $\GeV$ for each event. The background probability for each event is calculated with a $W$ + jets matrix
element from the {\sc{vecbos}} \cite{Vecbos:1991} generator.  Since there is no $\mtop$ dependence in the background probability, it is calculated only once for each event. 

\begin{figure}[htbp]
	\begin{center}
	\includegraphics[width=6cm]{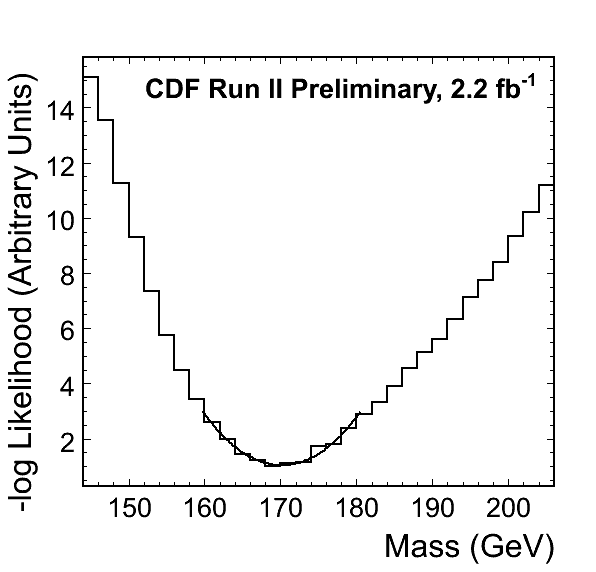}  
	\caption{Negative log likelihood (from Eq. \ref{eq:masslike}) as a function of top quark mass for all data events. The calibration functions have not yet been applied. \label{fig:massLike}}
	\end{center}
	\end{figure}

To improve the statistical uncertainty on the $\mtop$ measurement, we add a Gaussian constraint on the background fraction $\left( 1-c_s
\right)$ to the likelihood function.  The background fraction is constrained to be $0.498 \pm 0.106$ from Tab. \ref{tab:tau_m2}.  The
likelihood function is calculated as:
\begin{equation}
\mathcal{L} = \prod_{i=1}^NP\cdot\exp\left(-\frac{1}{2}\left(\frac{(1-c_s) - 0.498}{0.106}\right)^2\right), \label{eq:masslike}
\end{equation}	
\noindent where P is a combination of the signal ($P_{sig}$) and background ($P_{bkgd}$) probabilites with a relative normalization term ($A_{bkgd}$):
\begin{equation}
P = c_s P_{sig}\left(\vec{x}; M_{top}\right) + A_{bkgd}\left(1-c_s\right)P_{bkgd}\left(\vec{x}\right).
\end{equation}
\noindent The signal and background probability are both calculated by integrating over the differential cross section for the appropriate
process:
\begin{equation}
P_{sig/bkgd} = \frac{1}{\sigma_{sig/bkgd}}\int{d\sigma_{sig/bkgd}(\vec{y})f(\tilde{q_1})f(\tilde{q_2})W(\vec{x},\vec{y})d\tilde{q_1}d\tilde{q_2}},
\end{equation}
\noindent where $d\sigma$ is the differential cross section, $f$ is the parton distribution function (PDF) for a quark with momentum fraction of the incident proton $\tilde{q}$, $\vec{x}$ refers to detector measured quantities, $\vec{y}$ refers to parton level quantities, and $W(\vec{x},\vec{y})$ is the transfer function used to map $\vec{x}$ to $\vec{y}$.  After calculating the probabilities for each event, we evaluate a likelihood function for each of the 31 input top quark masses and fit the result with a second order polynomial to derive $\mtop$ and its statistical uncertainty.  We calibrate the measurement on 21 $\ttbar$ MC samples covering a mass range of 155 $\GeV$ to 195 $\GeV$. The likelihood function and fit for the data can be seen in Fig. \ref{fig:massLike}. We measure $\mtop$ to be $172.7 \pm 9.3 ~(\mathrm{stat.}) ~\GeV$.

\subsection{\label{sec:massSyst} Top Quark Mass Measurement Systematic Uncertainties and Result}
We consider 12 different sources of systematic uncertainty for the $\mtop$ measurement.  The largest uncertainty is from the JES (mentioned in Sec. \ref{sec:xsecSyst}). For this uncertainty, we shift the jet energies up and down by each correction's uncertainty and sum in quadrature the systematic uncertainty measured from each correction.  Since the top quark mass is very sensitive to the energy of its daughter particles, the
JES uncertainty is the dominant uncertainty for this measurement.

We also consider systematic uncertainties from the differences in
parton showering models from different MC generators, ISR and FSR, and
color reconnection by performing the measurement with MC models which
account for each effect.  The background fraction uncertainty is
measured by re-performing the measurement with pseudoexperiments where the
fraction of the background contribution from the QCD multijets
background and each of the $W$ + jets backgrounds is shifted within its
uncertainty from Tab. \ref{tab:tau_m2}.  The uncertainty from each
background is then added in quadrature to measure the total background
fraction uncertainty.

To measure the uncertainties from PDFs, we measure the shift
from the different CTEQ6 eigenvector PDFs.  The
uncertainty from the fraction of gluon-gluon fusion produced $t\bar{t}$
events is evaluated by reweighting the MC events so that the percentage of $t\bar{t}$
events which result from gluon-gluon fusion is shifted from 5\% to
20\%.

The $b$-jet uncertainty accounts for different fragmentation models and semileptonic branching ratios for jets from $b$
quarks \cite{bjet}.  These uncertainties are added in quadrature to an uncertainty measured from shifting the energy scale of jets from $b$ quarks to get the total $b$-jet uncertainty.  We also account for shifts from the lepton energy scale by considering changes in the measurement from MC samples with shifted $\tau$ energy.

The pileup systematic uncertainty accounts for a known mismodeling in the luminosity profile of the MC.  To evaluate this, we measure the shift in the measurement from MC events which are reweighted to give the luminosity profile which is seen in the data.

Finally, we consider systematic uncertainties from our calibrations. We shift the calibration function within its uncertainty to measure the calibration systematic uncertainty.  Even after the calibration function is applied, we find a 0.14 $\GeV$ uncertainty on the fit of the mass residual (defined as the true mass substracted from the measured mass) across all 31 mass points.  Due to this, we take a 0.14 \GeV systematic uncertainty for MC statistics.

The full table of systematic uncertainties for the top quark mass measurement can be found in Tab. \ref{tb:systtau}.

	\begin{table}
	\begin{center}
	\begin{tabular}{|l||r|}
	\hline
	Source  & Uncertainty ($\GeV$)\\
	\hline
	JES          & 3.37  \\
	MC Generator & 0.50  \\
	ISR/FSR      & 0.34  \\
	Color Reconnection & 0.50  \\
	Background Fraction   & 0.47  \\
	MC Statistics & 0.14 \\
	PDF           & 0.12  \\
	gg fusion     & 0.17  \\
	B-jet         & 0.39  \\
  Lepton $p_T$  & 0.19  \\
	Pileup        & 0.95  \\
	Calibration   & 0.17 \\
	\hline
	Total  &  3.7   \\
	\hline
	\end{tabular}
	\caption{Total systematic uncertainties on $M_{top}$ for the $\tau$ + jets channel.}
	\label{tb:systtau}
	\end{center}
	\end{table}	

Having evaluated all uncertainties, we measure $\mtop$ to be $172.7 \pm 9.3 ~(\mathrm{stat.})~ \pm 3.7 ~(\mathrm{syst.})~\GeV$ which agrees with the most recent Tevatron combination of $173.2 \pm 0.9 ~\GeV$ \cite{WAmass}.

\section{Conclusion}
We use the $\tau$ + jets decay channel to identify $t\bar{t}$ events as well as measure the top quark properties with 2.2 $\fb$ of data. We find the $t\bar{t}$ pair production cross section to be $8.8 \pm 3.3 ~(\mathrm{stat.})\pm 2.2 ~(\mathrm{syst.+lumi.})$ pb.  We also measure the top quark mass in this decay channel for the first time ever and find it to be $172.7 \pm 9.3 ~(\mathrm{stat.}) \pm 3.7 ~(\mathrm{syst.}) ~\GeV$. We find the measurements to be consistent with the current CDF combination top pair production cross section of $7.5 \pm 0.5$ pb \cite{CDFxsec} and the Summer 2011 Tevatron top quark mass combination of $173.2 \pm 0.9 ~\GeV$ \cite{WAmass}.  The values we measure with $\tau$ leptons agree with current measurements, therefore, we find no evidence against lepton universality.  Additionally, the success of these measurements demonstrates that we can do complicated analyses with $\tau$ leptons in high jet multiplicity environments at hadron colliders.

\begin{acknowledgments}
We thank the Fermilab staff and the technical staffs of the
participating institutions for their vital contributions. This
work was supported by the U.S. Department of Energy and National
Science Foundation; the Italian Istituto Nazionale di Fisica
Nucleare; the Ministry of Education, Culture, Sports, Science and
Technology of Japan; the Natural Sciences and Engineering Research
Council of Canada; the National Science Council of the Republic of
China; the Swiss National Science Foundation; the A.P. Sloan
Foundation; the Bundesministerium fuer Bildung und Forschung,
Germany; the Korean Science and Engineering Foundation and the
Korean Research Foundation; the Particle Physics and Astronomy
Research Council and the Royal Society, UK; the Russian Foundation
for Basic Research; the Comision Interministerial de Ciencia y
Tecnologia, Spain; and in part by the European Community's Human
Potential Programme under contract HPRN-CT-20002, Probe for New
Physics.
\end{acknowledgments}
\bigskip 

\begin{thebibliography}{99} 
\bibitem{cdfdet} D. Acosta {\it et al.}, Phys. Rev. D {\bf 71}, 032001 (2005).
\bibitem{tauSel} CDF Collaboration, /CDF/PUB/EXOTIC/PUBLIC/8676
\bibitem{secvtx} T. Affolder {\it et al.}, Phys. Rev. D {\bf 64}, 032002 (2001).
\bibitem{pythia} T. Sjostrand {\it et al.}, Comput. Phys. Commun. {\bf 135}, 238 (2001).
\bibitem{tauola} Z.~Was,
  Nucl.\ Phys.\ Proc.\ Suppl.\  {\bf 98}, 96-102 (2001).
  [hep-ph/0011305].
\bibitem{Aaltonen:2010ic}  T.~Aaltonen {\it et al.},
  Phys.\ Rev.\ Lett.\  {\bf 105}, 012001 (2010).
\bibitem{jes} A. Bhatti {\it et al.}, Nucl. Instrum. Meth., A {\bf 566}, 375 (2006).
\bibitem{ifsr} A. Abulencia {\it et al.}, Phys. Rev., D {\bf73}, 032003 (2006).
\bibitem{cr} D. Wicke and P. Z. Skands, Eur. Phys., J {\bf C52}, 133 (2007).
\bibitem{herwig} G. Corcella {\it et al.}, J. High Energy Phys., {\bf 01}, 010 (2001).
\bibitem{cteq6} J. Pumplin {\it et al.}, J. High Energy Phys. 07, {\bf 012}, (2002).
\bibitem{tagSF} J. Freeman, S. Grinstein, T. Junk, E. Palencia, CDF/DOC/SEC\_VTX/CDFR/9280
\bibitem{tauSF} Y. Tu and P. Murat, CDF/DOC/ELECTROWEAK/CDFR/8470
\bibitem{note9954} A. Mitra, S. Wang, and S. Tsai, CDF/ANAL/TRIGGER/CDFR/9954
\bibitem{bib:lumi} V.~Papadimitriou, eprint arXiv:1106.5182.
\bibitem{Abulencia:2007br}
A.~Abulencia {\it et al.},
Phys.\ Rev.\ Lett.\  {\bf 99}, 182002 (2007)
\bibitem{Mahlon:1995zn}
G.~Mahlon and S.~J.~Parke,
Phys.\ Rev.\  D {\bf 53}, 4886 (1996)
\bibitem{Vecbos:1991}
F.A.~Berends, H.~Kuijf, B.~Tausk and W.T.~Giele,
Nucl.\ Phys.\ B {\bf357}:32-64 (1991)
\bibitem{bjet} T. Aaltonen et al., Phys. Rev. D {\bf 79}, 092005 (2009).
\bibitem{CDFxsec} CDF Collaboration, /CDF/PUB/TOP/PUBLIC/9913
\bibitem{WAmass} CDF and D0 Collaborations, arXiv:1107.5255 [hep-ex].
\end{thebibliography}

\end{document}